\documentclass[journal]{IEEEtran}
\usepackage{amsmath,amssymb,graphicx}
\usepackage{authblk}
\usepackage{wrapfig}
\usepackage{booktabs}
\usepackage{microtype}
\usepackage{caption}
\usepackage[separate-uncertainty=true, multi-part-units=single]{siunitx}
\usepackage{bm}
\usepackage{enumitem}
\usepackage{float}
\usepackage[hidelinks]{hyperref}
\usepackage{glossaries}
\usepackage{etoolbox}
\newcolumntype{B}{<{\hspace{-1ex}}c}
\usepackage{multirow}
\DeclareSIUnit\px{px}
\DeclareSIUnit\db{dB}
\usepackage[utf8]{inputenc}
\usepackage[style=ieee,backend=biber, url=false,doi=false,isbn=false,eprint=false, maxbibnames=3,minbibnames=1]{biblatex}
\AtEveryBibitem{%
	\clearfield{day}%
	\clearfield{month}%
	\clearfield{endday}%
	\clearfield{endmonth}%
}

\newcommand{\ignore}[1]{\relax}
\expandafter\def\expandafter\normalsize\expandafter{%
	\normalsize
	\setlength\abovedisplayskip{6pt}
	\setlength\belowdisplayskip{6pt}
	\setlength\abovedisplayshortskip{6pt}
	\setlength\belowdisplayshortskip{6pt}
}

\newacronym{psf}{PSF}{Point Spread Function}
\newacronym{snr}{SNR}{Signal-to-Noise Ratio}
\newacronym{gpu}{GPU}{graphics processing unit}
\newacronym{bd}{BD}{Blind Deconvolution}
\newacronym{cnn}{CNN}{Convolutional Neural Network}
\newacronym{dnn}{DNN}{deep neural network}
\newacronym{map}{MAP}{Maximum a Posteriori}
\newacronym{tv}{TV}{Total Variation}
\newacronym{tvrl}{TV-RL}{Total Variation regularized Richardson-Lucy}
\newacronym{rl}{RL}{Richardson-Lucy}
\newacronym{ssim}{SSIM}{Structural Similarity}
\newacronym[plural=AFs,firstplural=Autofocuses (AFs)]{af}{AF}{Autofocus}
\newacronym{sml}{SML}{Sum of Modified Laplacian}
\newacronym{lapv}{LAPV}{Variance of Laplacian}
\newacronym{ewc}{EWC}{Energy of Wavelet Coefficients}
\newacronym{ws}{WS}{Wavelet Sparsity}
\newacronym{gss}{GSS}{Golden Section Search}
\newacronym{sd}{SD}{Standard Deviation}
\newacronym{dof}{DOF}{depth-of-field}
\newacronym{hpf}{HPF}{high-pass filter}
\newacronym{na}{NA}{Numerical Aperture}
\newacronym{roi}{ROI}{Region of Interest}
\newacronym{fov}{FOV}{field-of-view}
\newacronym{fwhm}{FWHM}{full width at half maximum}
\newacronym{gan}{GAN}{generative adversarial network}
\newacronym{sted}{STED}{stimulated emission depletion}
\newacronym{svm}{SVM}{support vector machines}
\newacronym{mri}{MRI}{magnetic resonance imaging}
\title{Free annotated data for deep learning in microscopy? A hitchhiker's guide}
\author{Adrian Shajkofci$^{1,2}$, Michael Liebling$^{1,3}$\thanks{$^{1}$Idiap Research Institute, CH-1920 Martigny, Switzerland\\
		$^{2}$\'Ecole Polytechnique F\'ed\'erale de Lausanne, CH-1015 Lausanne, Switzerland\\
		$^{3}$Electrical \& Computer Engineering, University of California, Santa Barbara, CA 93106, USA. \\SNSF Grants 206021\_164022, 200020\_179217. \\Published in Photoniques 104.}}

\begin{document}
%

\maketitle


\begin{abstract}

\vspace{5pt}
In microscopy, the time burden and cost of acquiring and annotating large datasets that many deep learning models take as a prerequisite, often appears to make these methods impractical. Can this requirement for annotated data be relaxed? Is it possible to borrow the knowledge gathered from datasets in other application fields and leverage it for microscopy? Here, we aim to provide an overview of methods that have recently emerged to successfully train learning-based methods in bio-microscopy.

\end{abstract}
%
%
\section{Introduction}
\label{sec:intro}
\begin{figure}
	\centering
	\includegraphics[width=0.65\linewidth]{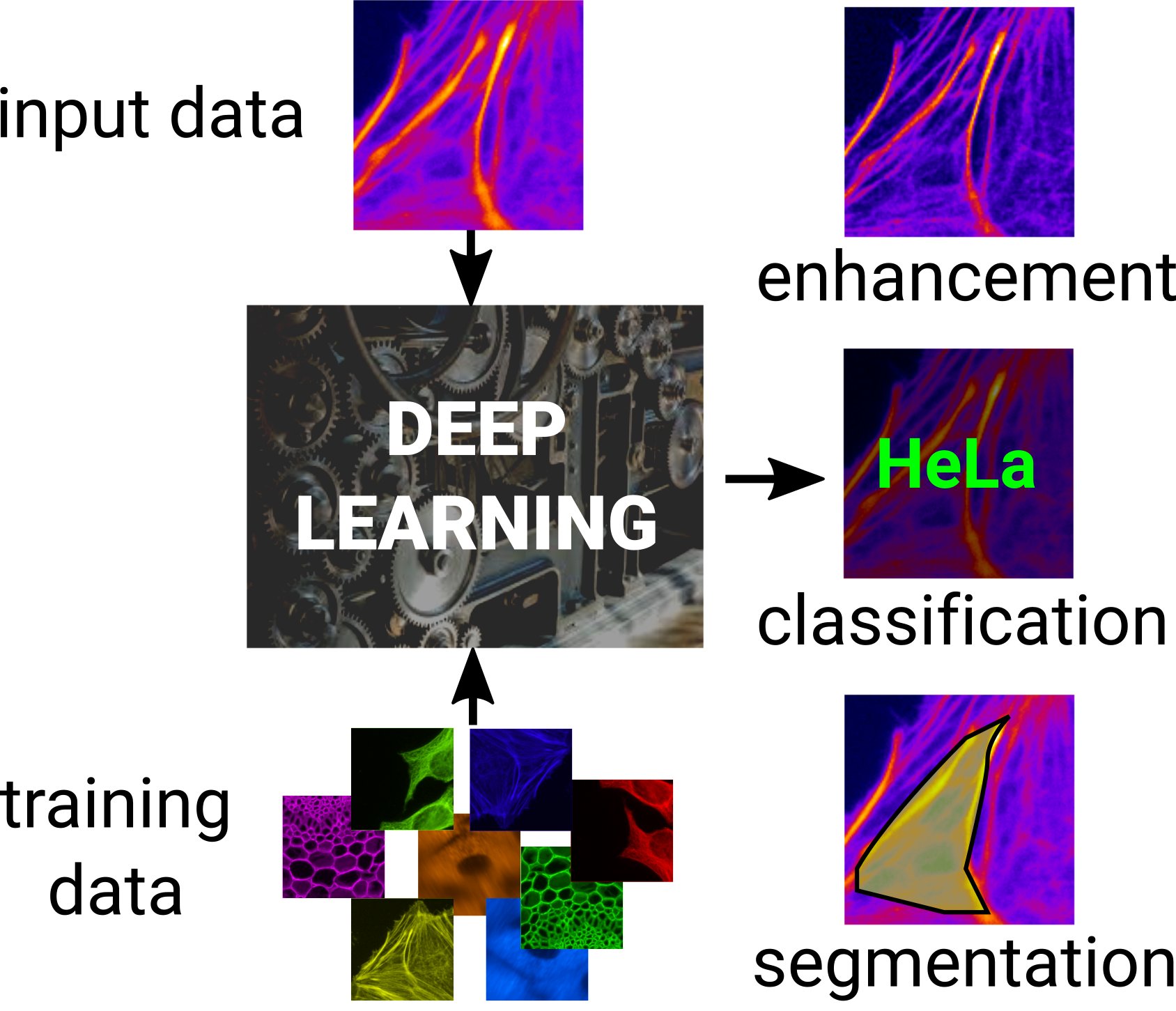}
	\caption{Deep learning in microscopy: overview.}
\end{figure}
Optical microscopy, despite being an invaluable tool in biology and medicine to observe and quantify cellular function, organ development, or disease mechanisms, requires constant trade-offs between spatial, temporal, and spectral resolution, invasiveness, acquisition time, and post-processing effort. As for other imaging fields, learning-based techniques are having a major impact in microscopy, where their potential to improve resolution, reduce invasiveness, or increase the speed of microscopy acquisitions has recently been demonstrated \cite{belthangady_applications_2019}. These techniques, in particular the ones that involve \glspl{dnn}, have benefited from the shift of intensive computing tasks to \glspl{gpu} that has taken place over the last decade.
Yet despite the ever-increasing computational power, it is often the lack of labeled training data that is the limiting factor for wide adoption. Annotating data is often a lengthy and expensive task, since it involves tedious work, generally by skilled experts. Annotation can be especially challenging in the case of three-dimensional images, common in microscopy, despite the development tools dedicated to this task \cite{yushkevich_user-guided_2006,grunberg_annotating_2017}. While the acquisition and annotation of volumetric data is common in some medical settings and for certain modalities because of their wide use for healthcare applications (e.g. \gls{mri}, annotated by trained radiologists), bio-microscopy applications often lack similarly large and high-quality annotated volumetric datasets.

\begin{figure*}
	\centering
	\includegraphics[width=0.75\linewidth]{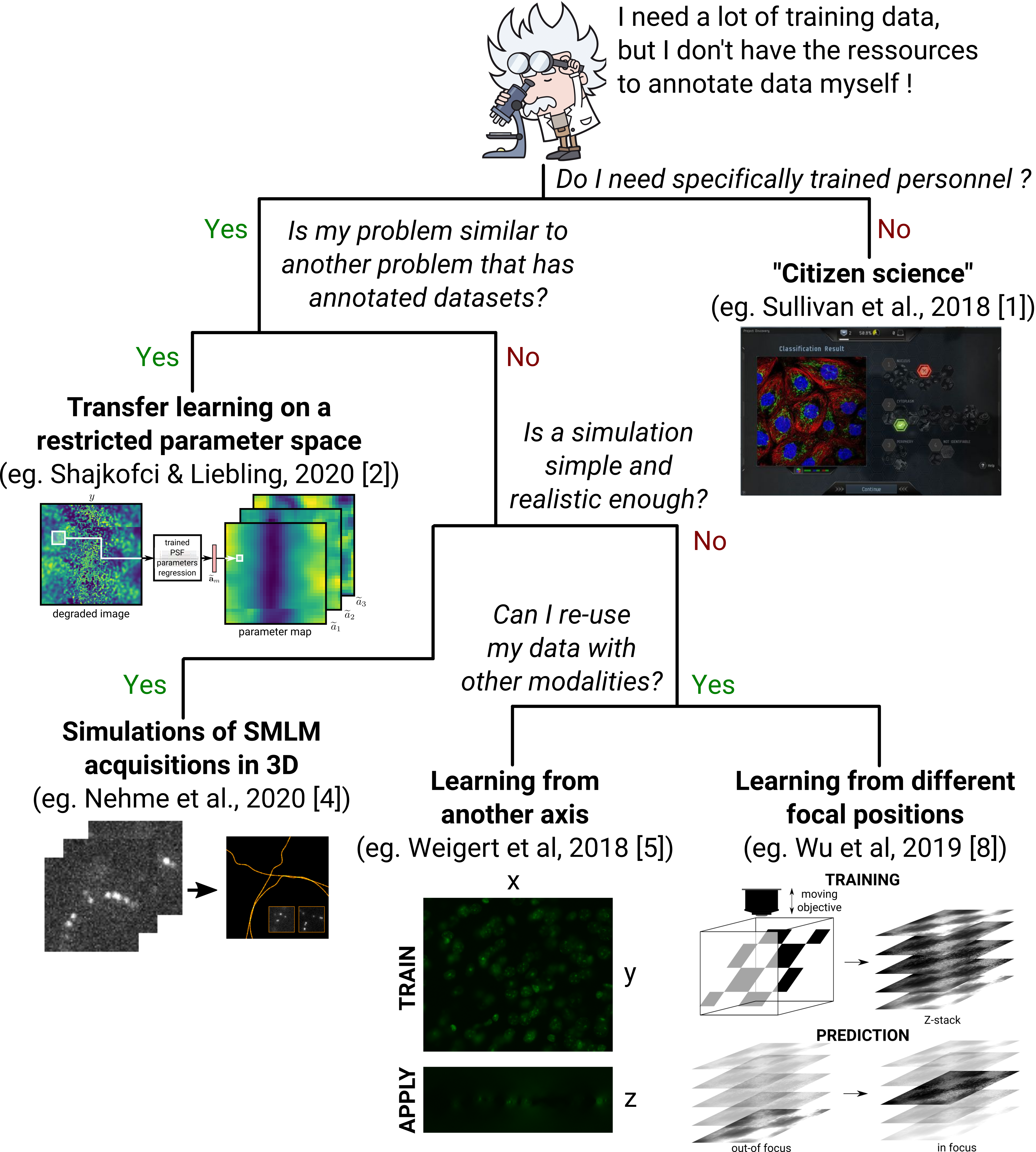}
	\caption{The quest for less painful annotation: overview of questions and methods. The illustrations were made by AS and adapted from the original papers.}
\label{fig:quest}
\end{figure*}

For many applications in biology, the time burden, cost, or physical feasibility of acquiring and annotating datasets for deep learning models \emph{de novo} is simply out of the question. Can this requirement for annotated data be relaxed? Is it possible to borrow the knowledge gathered from datasets in other application fields, such as e-commerce or computer gaming, and leverage it for bio-microscopy? Specifically, could annotated datasets be generated from realistic synthetic models of tridimensional objects? Or could more abstract prior knowledge about the data be used to enhance the resolution?
Here, we aim to provide an overview of some solutions that have emerged to tackle the problem of gathering sufficient annotated data to train learning-based methods in bio-microscopy. We have grouped the approaches in four broad categories: developing manual annotation strategies, learning from annotated images from other domains, building annotations from simulations, and using self-annotated data. This quest for annotated data is summarized in Figure~\ref{fig:quest}.

\section{Overview of Data Annotation Strategies for Deep Learning methods in Biomicroscopy}
\subsection{Learning from large training datasets}
Image segmentation is a common computer vision task in which each pixel of an image is assigned a label corresponding to the object it belongs to. In microscopy, it is used, for example, to delineate, identify, and count cells. Learning-based segmentation methods require a training data set composed of images together with masks that correspond to the objects and their coordinates in the  images. This annotation is usually performed manually and can be tedious. Other fields have integrated the task of annotating images into security forms on the web (to exclude robots from accessing content) or into entertaining game puzzles, such as to entice the public to provide quality annotations. 
Sullivan et al. have proposed to annotate microscopy data by a similar approach, as part of a multiplayer computer game named Eve Online \cite{sullivan_deep_2018}. Using the publicly available data set from the Cell Atlas of the Human Protein Atlas (HPA), they obtained, over one year, nearly 33 million classifications of subcellular localization patterns of immunostained proteins in 20 different organelles and cellular structures. The results were successfully used as a training dataset for a segmentation \gls{dnn}.

\subsection{Learning from other domains}

Training a machine learning algorithm by re-using computer vision data sets that were originally intended for other tasks is at the core of transfer learning. This approach can overcome the lack of annotated data in one field (such as microscopy) if annotated data exists in a different field. A wide variety of annotated image datasets are available, such as ImageNet \cite{krizhevsky_imagenet_2012}, MS-COCO \cite{lin_microsoft_2014-1} or Places \cite{zhou_places_2018}, which contain foremost scenes depicting everyday objects and situations. In addition to allowing access to a large number of examples,  learning from natural-scene images can help avoid learning on images that contain unwanted aberrations, such as  blur and low-light noise commonly found in microscopy images. It becomes essential when developing methods that aim at removing aberrations, for example when using a \gls{dnn} pipeline for deconvolution or denoising of microscopy images, since the ground truth of natural-scene photographs is more readily accessible that fine structures that challenge the resolution limits of the most powerful microscopes\ignore{\cite{yuanhao_gong_natural-scene_2016}}.

In most cases of transfer learning, pre-trained models are used as feature extractors \cite{kim_transfer_2017}, as the input of an unsupervised classifier such as \gls{svm}, or are fine-tuned with job-specific training data. However, there are some applications where transfer learning can be used without fine-tuning, specifically, when the feature space can be mapped identically in both domains.
For image deconvolution and depth estimation, a description of the  \gls{psf} is a pre-requisite and accurately determining \gls{psf} parameters is therefore essential.
Recently, we formulated the problem of retrieving the physical \gls{psf} parameters of an optical microscope as a regression task for which we trained a \gls{dnn}. Interestingly, when learning from  data that consist of textured images, even if they are different from the end-use application (and possibly unrelated to microscopy), the trained model remains just as accurate as when it is trained on microscopy images \cite{shajkofci_spatially-variant_2020-1}. The ability to train on generic data also helps prevent over-fitting the trained model to a specialized and narrow data set, which would make the model less suitable to be used in other situations. It also suggests the possibility of generalizing the trained network to other data types, provided that the new data and the data used for training share a common feature space.

\subsection{Learning from simulated data sets}

Generating annotated data from simulations is another effective approach to produce large and reliable training data sets. In some computer vision applications, such as autonomous car driving, data from simulated computer graphics 3D environments have been effectively used to train segmentation methods. Examples of such datasets include the Flying chairs dataset, or the GTA5 dataset derived from a computer game. Generally, the accuracy of \gls{dnn} trained only on simulated data is poor, due to the extreme complexity of natural scenes that can hardly be reproduced with simple simulations.

In microscopy, image complexity remains fairly low in some modalities such as single-molecule localization microscopy (SMLM), where features consist of dots. The image processing task, for which \gls{dnn} have been used, consists in converting images of random subsets of activated fluorophores, obtained over many consecutive diffraction-limited frames, into a high-precision point cloud.
Data simulation of a realistic diffraction-limited ground truth is achievable, for example,  by filtering the expected (punctate) objects by the optical \gls{psf} and take into account a realistic noise model \cite{sage_super-resolution_2019}. Such approaches have allowed recent \gls{dnn} methods such as DeepStorm3D \cite{nehme_deepstorm3d_2020} to recover densely overlapping \glspl{psf} of multiple emitters over a large axial range and output a list of their 3D positions. The corresponding training data set were created by simulating a large number of images using randomly generated 3D patterns and the phase mask that governs the \glspl{psf} modelled on the physical implementation of the microscope. In order for the simulations to work as a sole training data, one possible way is to restrict the parameter space (i.e. the output of the network) by closely modeling the data generation model to the optics or choose a subset of plausible physical parameters (\cite{shajkofci_spatially-variant_2020-1, nehme_deepstorm3d_2020}).

\subsection{Learning from the input image itself}

\glspl{dnn} are also used for image enhancement such as deblurring (going from a blurry image to a sharper image), denoising (going from a noisy to a clean image), and for super-resolution (going from a low-resolution image to a higher-resolution image). These networks are traditionally trained on pairs of clean and distorted images. In microscopy, where the raw images often already reach the physical limits due to diffraction, a higher-resolution ground-truth image is simply not accessible. To cope with this problem, it is sometimes possible to use  approaches that exploit particular features of the data, such as its isotropy or the availability of complementary imaging modalities.

Weigert et al. \cite{weigert_content-aware_2018} proposed a pipeline aimed at restoring images using semi-synthetic training data. Specifically, it restores the axial resolution of volumetric images  lost due to the axial elongation of the optical \gls{psf} and the low axial sampling rate. By assuming that similar features are to be expected regardless of sample orientation, the method leverages the fact that these features can be much better resolved in lateral views than in depth, hence the training to improve depth-resolution is done based on the latter images.
Nevertheless, for many other applications, access to higher quality images or synthetic data is not available. To overcome this limitation, Krull et al. \cite{krull_probabilistic_2020} developed a self-supervised training method for denoising based on the assumption that the noise is independent from pixel to pixel and that the true intensity of a pixel can be predicted from the local image context since it is not locally independent. The method involves a noise model whose probability distribution is learned from the training data and  a network is trained to discriminate the underlying image from the noise. As a training set, a very small dataset of noisy images of the same type can be used. This method is virtually equivalent to training a \gls{dnn} for every specific noise distribution.

A similar training scheme was also used in \cite{wang_deep_2019, wu_three-dimensional_2019}, where the authors trained a \gls{gan}-based \gls{dnn} to transform an acquired low-resolution image into a high-resolution image using matched pairs of experimentally acquired images after registration and alignment. In \cite{wang_deep_2019}, these pairs came from images of the same object using a confocal microscope and a super-resolved \gls{sted} microscope. In \cite{wu_three-dimensional_2019}, the authors used a similar \gls{dnn} to generate images that look as if they had been taken from another focal plane by training from images acquired at different heights in the sample.
For both applications, the authors caution that the network must be (re)-trained for each specific image modality or experimental setup, as the methods do not produce ideal results otherwise. The application of such methods therefore remains somewhat limited to cases where the type of images and microscope settings are known beforehand and where a high number of similar images can be acquired, which could be particularly relevant for time-lapse imaging, high-resolution 3D stacks, or imaging of histological samples prepared under controlled and standardized conditions.

\section{Conclusions}
Deep learning technologies have enabled multiple applications that are transforming our day-to-day routines, including the way we approach microscopy. 
While limitations such as network capacity (can networks learn to predict from the wide variety of data types common in microscopy?), generalization (can a network trained on one type of data be used to handle other types of data?), and overfitting (is the network limited to predicting only what it has already seen?) are some pressing issues that the field is facing,  
the lack of good quality training data is likely the single most important aspect that affects accuracy and effectiveness of tasks such as enhancement, classification or segmentation.
The most promising methods to overcome the scarcity of training data appear to leverage prior knowledge of the physical objects and image formation process \cite{weigert_content-aware_2018, nehme_deepstorm3d_2020}, \cite{shajkofci_spatially-variant_2020-1}, or of the noise distribution \cite{krull_probabilistic_2020}.

Even if we are still a long way from a blind pipeline that will enhance, classify or segment biological data without tedious annotation work, good knowledge of the problem and assumptions about the data allow scientist to already reap the benefits of deep learning tools by crafting adapted training sets without having to produce or wait for the availability of large annotated sets.

\section{References}
\label{sec:references}
	\printbibliography[heading=none]

\end{document}